\documentclass[twocolumn]{aastex631}

\usepackage{soul}
\usepackage{bm}
\usepackage{amsmath}
\begin{document}

\title{Exploring variation of double-peak broad-line profile in strongly variable AGNs}
\author[0000-0002-2581-8154]{Jiancheng Wu}
\affiliation{Department of Astronomy, School of Physics, Huazhong University of Science and Technology, Luoyu Road 1037, Wuhan, China}
\author[0000-0003-4773-4987]{Qingwen Wu$^*$}
\affiliation{Department of Astronomy, School of Physics, Huazhong University of Science and Technology, Luoyu Road 1037, Wuhan, China}
\author[0000-0002-2310-0982]{Kaixing Lu}
\affiliation{Yunnan Observatories, Chinese Academy of Sciences, Kunming 650011, People’s Republic of China}
\affiliation{Key Laboratory for the Structure and Evolution of Celestial Objects, Chinese Academy of Sciences, Kunming 650011, People’s Republic of China}
\author[0000-0002-2355-3498]{Xinwu Cao}
\affiliation{Institute for Astronomy, School of Physics, Zhejiang University, 866 Yuhangtang Road, Hangzhou 310058, People’s Republic of China}
\author[0009-0003-0516-5074]{Xiangli Lei}
\affiliation{Department of Astronomy, School of Physics, Huazhong University of Science and Technology, Luoyu Road 1037, Wuhan, China}
\author[0000-0001-5019-4729]{Mengye Wang}
\affiliation{Department of Astronomy, School of Physics, Huazhong University of Science and Technology, Luoyu Road 1037, Wuhan, China}
\author[0000-0001-7350-8380]{Xiao Fan}
\affiliation{Department of Astronomy, School of Physics, Huazhong University of Science and Technology, Luoyu Road 1037, Wuhan, China}

\begin{abstract}
The geometry and kinematics of the broad-line region (BLR) in AGNs are still unclear, which is crucial for studying the physics and evolution of supermassive black holes (SMBHs) and AGNs. The broad-line profile provides valuable information on BLR geometry and kinematics. In this work, we explore the evolution of line profiles in variable AGNs based on the BLR model of Czerny \& Hryniewicz, where the BLR is driven by the radiation pressure acting on dust at the surface layers of the accretion disk. The line profiles in the low-Eddington-ratio regime show a double-peak profile, which will become a single peak at high Eddington ratios. The high metallicity of $Z\gtrsim 5Z_{\odot}$ is required to reproduce the observational anti-correlation between the peak separation of broad lines and the Eddington ratio for a sample of AGNs. For the broad lines in variable AGNs, it will take several years to several decades to change their line profile if the disk luminosity suffers strong variation in a much shorter timescale. More monitoring of the broad line and continuum in strongly variable AGNs can shed special light on BLR physics.
\end{abstract}


\section{Introduction} \label{sec:intro}
Active Galactic Nuclei (AGNs) are one of the most luminous astrophysical objects in the universe, which are believed to be powered by the accretion of gas onto the central supermassive black holes \citep[SMBHs, e.g.,][]{Shakura1973}. Among the multi-component structure of AGNs, the broad lines with a velocity of several thousand kilometers per second originate from the pc-scale broad-line region (BLR), where the high-velocity gas moves surrounding the central SMBH. The UV photons from the accretion disk ionize the BLR gas and produce strong lines. The unified model of AGNs tried to explain the observed diversity in AGN properties by considering the orientation of the observer relative to the accretion disk and the surrounding torus of gas and dust, where the broad lines are observed in type I AGNs but are obscured in type II sources \citep[e.g.,][]{Antonucci1993, Urry1995}. Most AGNs show some level of intrinsic variability, which can be used to constrain the structure of these objects \citep[e.g., disk and emission line properties, e.g.,][and references therein]{Grier2013}. In particular, a growing number of sources have shown dramatic flux and spectral changes in both the X-rays and the optical/UV bands in the last two years, which are commonly called changing-look AGNs \citep{Tohline1976, Denney2014, Ricci2023}. The appearances or disappearances of the broad lines along with the variation of the continua provide a chance to explore either the evolution of the BLR or the variation of column density for X-ray absorption \citep[e.g.,][]{McElroy2016, Storchi2017, Lyu2021}.

In the AGN unification scheme, the BLR is located between accretion disk and dust torus based on the reverberation mapping (RM) observations \citep[e.g.,][]{Antonucci1993, Peterson1993, Wang2014, Netzer2015, Wangjm2022}. However, the origin of the BLR clouds is still unclear \citep[e.g.,][]{Sulentic2000, Peterson2004, Kaspi2005}. Theoretical models of BLR formation can generally be divided into two categories, where the BLR clouds either originate from gas blown off from the accretion disk \citep[e.g.,][]{Blandford1982, Begelman1983, Murray1995, Czerny2011}, or the clouds captured from outside \citep[e.g.,][]{Done1996, Hu2008, Wang2017}. The RM method based on the delayed response of BLR clouds to variations of the AGN continuum provides us a precise radius of BLRs \citep{Blandford1982, Peterson1993}. The RM measurements provide an empirical correlation between the time lag (or BLR size) and $5100{\text \AA}$ continuum luminosity \citep{Kaspi2000, Bentz2013}, even though the super-Eddington AGNs seem to deviate from this correlation \citep[][]{Du2019}. Information on the geometry and kinematics of BLR can also be obtained from the broad-emission line profile, which is always complex. Apart from the single-peaked profile, some of the double-peaked broad lines are also reported \citep{Chen1989, Eracleous1994}. In the Solan Digital Sky Survey (SDSS), it is found that a few percent of AGNs show double-peaked broad lines \citep[e.g.,][]{Strateva2003}.  The velocity field of BLR may be a superposition of different components, such as rotation, turbulence, shock component, inflow/outflow components, and Doppler motion, where the different components will lead to different line profiles. The single-peaked profile can originate from explained either disk-like or spherical distribution of BLR clouds. The direct resolved imaging of 3C 273 using the Very Large Telescope Interferometer (VLTI) supports the disk-like BLR model \citep{Gravity2018, Gravity2021}. For double-peaked lines, it can be well fitted by the emission from photoionized gas in a circular or elliptical disk around a central black hole \citep{Chen1989, Eracleous2003}. It should be noted that alternative models have also been proposed for the origin of the double-peaked line emission, including bipolar outflows, binary black holes with separate BLRs, and an anisotropically illuminated spherical distribution of clouds \citep{Eracleous1995, Storchi2003, Cao2006, Flohic2012, Du2023}.

In the time-domain era, more and more AGNs show strong variability within a couple of years, which provides a valuable chance to explore the BLR physics based on their evolution. In this work, we explore the variation of the broad-line profiles from the BLR kinematics based on a model of radiatively driven dusty outflow. In Section \ref{model}, we describe our BLR model. The simulation results are presented in Section \ref{result}. Discussion and conclusion are in Section \ref{discussion}.

\section{BLR model}\label{model}
\subsection{Geometry and Kinematics}
In this work, we explore the structure and kinematics of BLR based on a physically motivated BLR model of \cite{Czerny2011}, where the failed radiately accelerated dusty outflow (FRADO) is considered as the physical mechanism for the formation of the BLR gas. The FRADO model can well explain the line profile and the observed BLR radius-luminosity correlation \citep[e.g., ][]{Czerny2011, Czerny2017, Naddaf2021, Naddaf2022, Naddaf2023}. In our simulations, the dynamics of clouds are the same as that of \cite{Naddaf2021}. We simply summarize the FRADO model as follows.

\begin{figure}[t]
\includegraphics[scale=0.55]{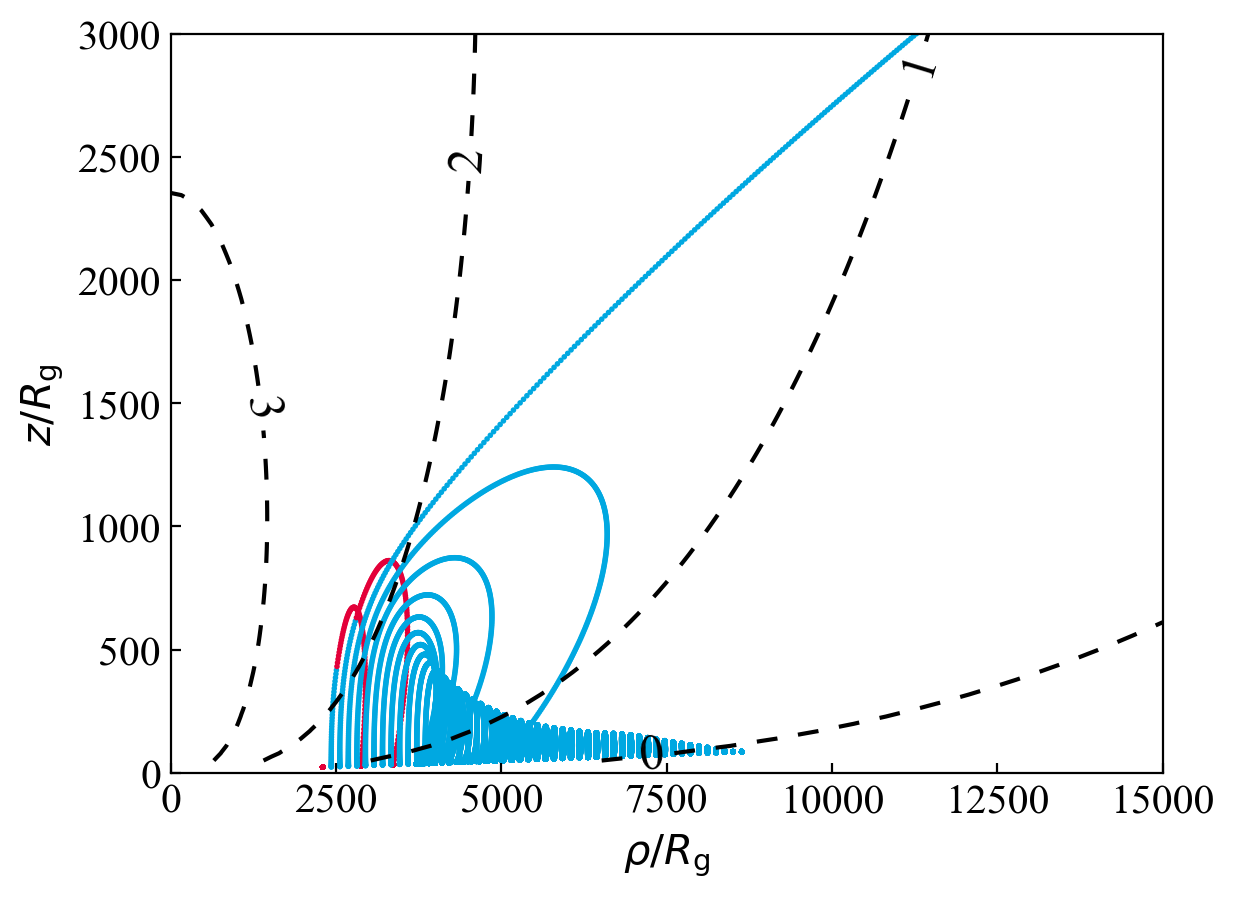}
\caption{An example of the BLR cloud trajectory for $M_{\rm BH}=10^8M_{\odot}$ and $\dot{m}=0.1$ in $\rho-z$ plane, where $\rho$ is the distance to central SMBH along disk, $z$ is the hight from the disk plane and the metallicity is set as $Z/Z_{\odot}=5$. The blue trajectory represents that the clouds embedded with dust grains are blown away by the radiation pressure. The red trajectory represents the clouds without dust grains, and the clouds undergo ballistic motion. The black dashed contours show the ratio of the local photoionization flux to the critical photoionization flux in log scale, $\log (\Phi/\Phi_{\rm c}$), where the critical flux $\Phi_{\rm c}$ corresponds to the photon flux that most effectively emits the lines.
\label{fig1}}
\end{figure}

The dust grains can survive in the accretion disk if the local disk effective temperature is lower than the dust grains sublimation temperature $T_{\rm s}\sim1500\,{\rm K}$. The local radiation pressure will push the dust grains and gas away from the disk surface, with the dust assumed to be strongly coupled with the gas in the FRADO model \citep{Czerny2011, Czerny2017}. We consider the 3D trajectory of gas clouds, where the motion is primarily controlled by the radiation pressure from the accretion disk and the gravitational field of the central SMBH. The net acceleration is $\bm{a}^{\rm net} = \bm{a}^{\rm gra} + \bm{a}^{\rm rad}$, where the radiative acceleration will play a role only if the dust temperature calculated in the cloud along the trajectory is less than the dust sublimation temperature. After the sublimation of the dust, the acceleration from the radiation pressure $\bm{a}^{\rm rad}=0$, and the ballistic motion will continue until the clump falls back onto the disk or escapes at a very large velocity. The computations of the dusty cloud motion, based on the determination of the radiation pressure from the extended accretion disk, can be found in the Appendix of \cite{Naddaf2021} for more details, where the opacity for gas and dust are carefully treated. To consider the gas/dust with different metallicities ($Z$), we adopt dust-to-gas ratio $\Psi=0.005 \times Z/Z_{\odot}$ \citep{Naddaf2022}, where $\Psi=0.005$ for solar abundance \citep{MRN}. The detailed content of the FRADO model can be found in \cite{Naddaf2021} and \cite{Naddaf2022}. According to radiatively driven wind model, the shielding effect for launching region should be important to avoid being over-heated \citep[e.g.,][]{Gallagher2007, Proga2007, Higginbottom2014}, where the shielding can be naturally caused by other clouds, wind, warm absorber or disk itself etc. \citep[e.g.,][]{Murray1995, Kartje1999}. We adopt their $\beta$-patch model in our simulations, where the outer part of the disk and the azimuthally extended region are more visible to the cloud than the $\alpha$-patch model. Simple geometrical models of the local shielding of the dusty cloud are considered following \cite{Naddaf2021}, which will not affect the line profile since no line emission is expected from the disk itself due to the shielding effect. The outer radius of the accretion disk is set to $R_{\rm out}=10^5\, R_{\rm g}$ ($R_{\rm g}$ is the gravitational radius), which is roughly consistent with the inner radius of the torus. 

To estimate the disk mass loss rate or the number of clouds from the disk surface at a given radius, we adopt $\dot{M}_{z} \propto \rho^{-2.5} \dot{M}$ as proposed by \cite{Czerny2017}. In some cases, the clouds are accelerated far away in the radial direction (e.g., see the trajectory in Figure \ref{fig1}). We simply assume that the contribution of the clouds is neglected if their radial distance is larger than the size of torus $R_{\rm torus}$, where the clouds may collide and merge with the torus. 

We show an example of the clump trajectories in Figure \ref{fig1}, where the clouds initially rise upward accompanied by radial outward motion and then fall back to the surface of the outer accretion disk. We assume that the dusty/gaseous clouds move as single entities along their respective trajectories. Assuming the clouds are continuously launched, the trajectories for all clouds can be assumed to the BLR geometry and the line luminosities can be derived by integrating contributions from all the clouds as distributed in Figure \ref{fig1}.

\begin{figure}[t]
\includegraphics[scale=0.6]{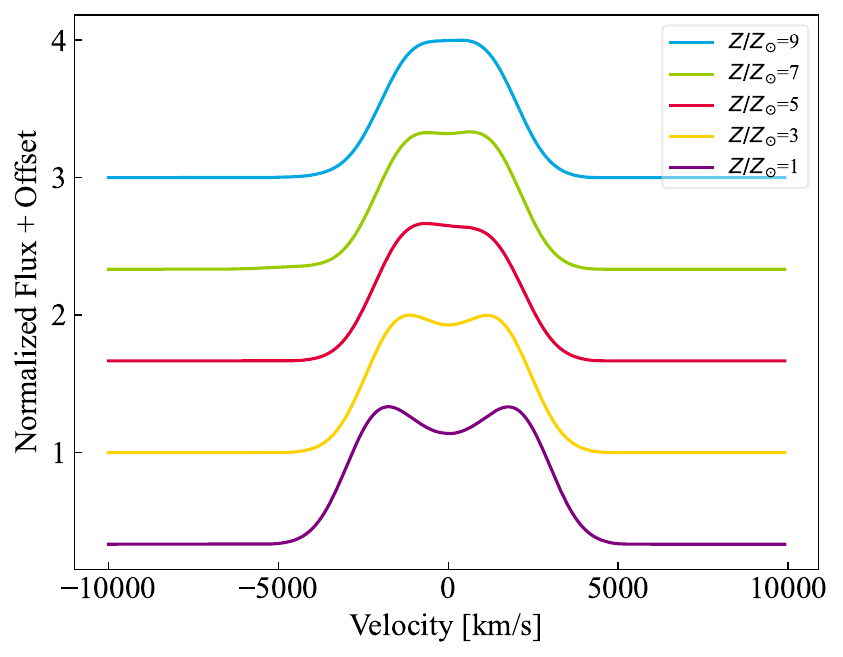}
\caption{The example of broad-line profile for different metallicities with $M_{\rm BH}=10^8M_{\odot}$, $\dot{m}=0.1$ and $i = 30^{\circ}$. 
\label{fig2}}
\end{figure}

\begin{figure*}[t]
\includegraphics[scale=0.75]{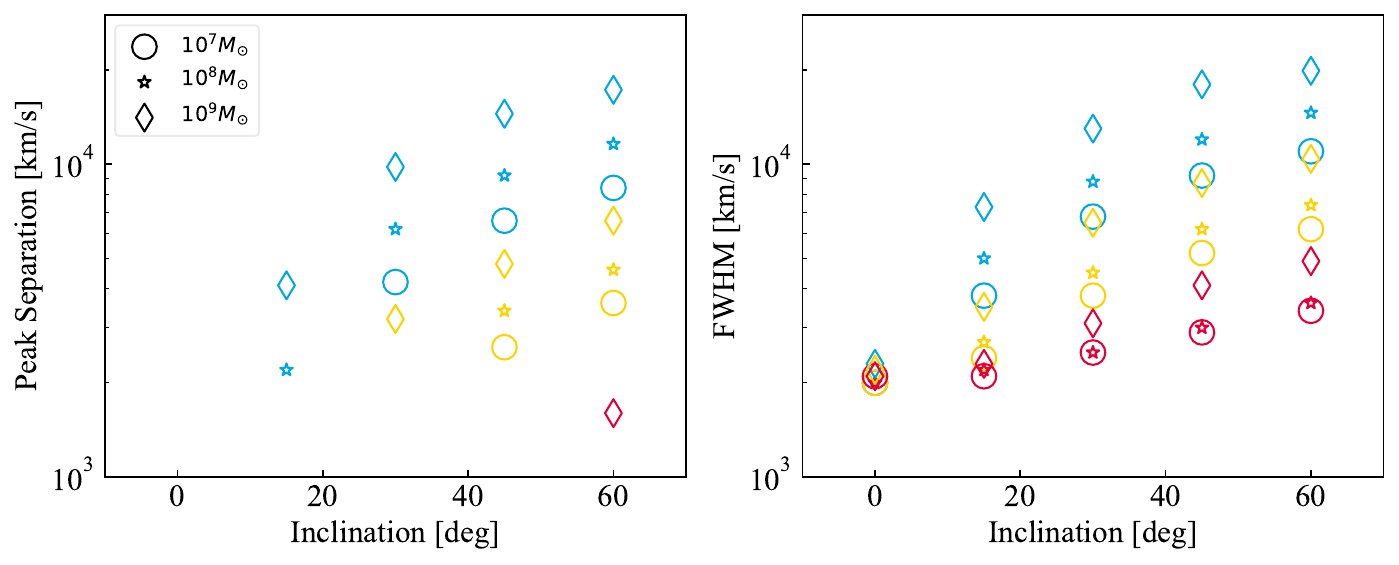}
\caption{The double-peak separation (left panel) and FWHM (right panel) of broad lines along with different inclination angles. The red, yellow, and blue symbols represent $\dot{m}=1, 0.1, 0.01$, respectively. The metallicity is fixed at $Z/Z_{\odot}=5$.
\label{fig3}}
\end{figure*}

\subsection{Clouds Emissivity and Line Profile}
The FRADO model presents the structure of clouds surrounding the SMBHs at a given dimensionless accretion rate. To calculate the line profiles, the emissivity of each cloud should be considered. Over the past decades, the photoionization model has successfully explained the emission line intensities in AGNs \citep[e.g.][]{Baldwin1995, Ferland2003, Korista2004}. In this work, the density of each BLR cloud during the motion is not calculated self-consistently. Consequently, we cannot directly calculate the line emission using the photoionization model. Instead, we adopt a broken power-law distribution of emissivity for each BLR cloud,

\begin{equation} \label{equ4}
    \epsilon = \left\{
    \begin{aligned}
        &\epsilon_{0}(\frac{\Phi}{\Phi_{\rm c}})^q, \qquad \Phi\leq\Phi_{\rm c}, \\
    &\epsilon_{0}(\frac{\Phi}{\Phi_{\rm c}})^{-q}, \qquad  \Phi>\Phi_{\rm c},
    \end{aligned}
    \right.
\end{equation} 
where $\epsilon_0$ is a constant, and $\Phi$ is the hydrogen-ionizing photon flux from the central accretion disk. The broken power-law is roughly consistent with the emissivity as adopted in modeling the double-peak lines in \cite{Chen1989} and also roughly consistent with the Cloudy simulation at given density \citep[see also,][]{Korista2004}. The parameter $q=1$ is adopted in our calculation. The adopted emissivity is different from that adopted in \cite{Naddaf2022}, where the emissivity is proportional to the altitude of the cloud.  The ionizing photon flux for a cloud can be described as,
\begin{equation} \label{equ5}
    \Phi = \int \frac{L_{\nu}}{2\pi r^2 h\nu}\,d\nu \times {\rm sin}\frac{z}{\rho},
\end{equation} 
where $L_{\nu}$ is the disk monochromatic luminosity, $r$ is the distance from the BH to the cloud, while $z$ and $\rho$ are the axial and radial positions of the cloud in cylindrical coordinates, respectively. The critical hydrogen-ionizing photon flux $\Phi_{\rm c} = 10^{18} \, {\rm cm^{-2}\, s^{-1}}$ is adopted, which corresponds to the maximum emissivities as constrained from the AGNs \cite[e.g.,][]{Korista2004}. The ionization emission is calculated from the global slim disk \citep{Abramowicz1988, Feng2019}, which will become the standard Shakura-Sunyaev disk at sub-Eddington rate \citep[e.g., $\dot{m}\lesssim 0.5$,][]{Shakura1973}. In this work, we focus on sub-Eddington accretion in most cases.

With the BLR structure and adopted emissivity, we can reproduce the line profiles based on BLR clouds distributed along the trajectories as shown in Figure \ref{fig1}. We assume the material is ejected from the disk and moves along the trajectory. We integrate the total line luminosity by considering each cloud distributed in the trajectories. Due to the limited number of clouds in our calculations, we consider a local turbulent broadening to smooth the line profile, where $\sigma = {\rm 850\, km\,s^{-1}}$ is adopted as constrained from fitting the disk-like H$\alpha$ profile of Arp 102B \citep[][]{Chen1989_2}. The value of turbulent velocity will slightly affect the line profile, however, the separation of two peaks or FWHM is roughly unchanged. Therefore, our conclusion will not be changed even though considering the uncertainties of the local turbulent broadening \citep{Rees1987, Bottorff2000_2}.

\section{Result}\label{result}
\subsection{The anti-correlation of double-peak separation and Eddington ratio: theories and observational constraints}
The profile of broad emission lines is strongly regulated by the mass of SMBH, accretion rate, the viewing angle, and gas metallicity \citep[see also,][]{Naddaf2022}. We further explore the variation of the line profiles and compare it with the observations. 

As an example, we present the line profiles in Figure \ref{fig2} for different metallicities from $Z=1Z_{\odot}$ to $Z=9Z_{\odot}$ for typical parameters of $M_{\rm BH}=10^8M_{\odot}$, $\dot{m}=0.1$ and inclination angle of $30^{\rm o}$. It can be found that the double-peak line profile will change to the flat-top profile with the increase of the metallicity. The clouds will suffer a stronger radiation-pressure force in the case of higher metallicity, which corresponds to a higher dust-to-gas ratio or higher total cross-section of dust grains. The BLR structure will be more disk-like at the lower metallicities, and therefore, the double-peak profile is more evident.

In Figure \ref{fig3}, we present the peak separation and FWHM of the broad lines in different inclination angles. For given other parameters, the peak separation or FWHM will increase with the increase of inclination, where the double peak is not evident at very low inclinations (e.g., $\sim 0^{\rm o}$). The BH mass also affects the line profile, where the higher BH masses will lead to larger peak separation or FWHM due to the BLR being closer to the SMBH in the case of higher mass. The physical reason is that the dust can survive at a smaller radius ($\rho=R/R_{\rm g}$) due to the lower disk temperature in the case of more massive black holes. The peak separation or FWHM  of broad lines will also increase at the lower accretion rate, where the disk temperature becomes lower and dust can survive at regions closer to SMBH. A lower inclination angle and higher accretion rate will easily lead to a single peak profile.

\begin{figure}[t]
\includegraphics[scale=0.7]{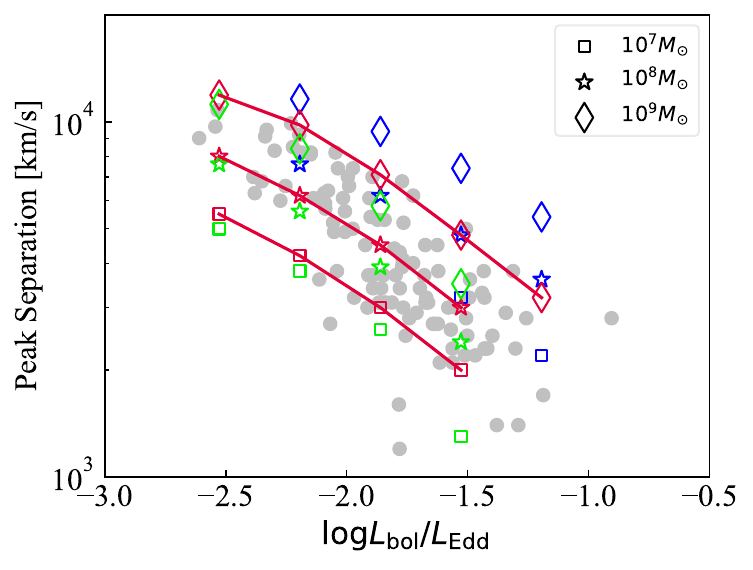}
\caption{The relation between the peak separation and the Eddington ratio, where the grey circles are the observational results for a sample AGNs \citep[][]{Wu2004}. The blue, red, and green represent the metallicity $Z/Z_{\odot}=1, 5, 9$ respectively. The bolometric luminosity $L_{\rm bol}$ in our model is derived from the Shakura-Sunyaev accretion disk. The inclination $i=30^{\circ}$ is adopted.
\label{fig4}}
\end{figure}

In Figure \ref{fig4}, we present the relation between the peak separation and bolometric Eddington ratio. They roughly follow an anti-correlation at given parameters, which is roughly consistent with the observational results of \cite{Wu2004}. Considering the BH mass is around $8<{\rm log}M/M_{\odot}<9.5$ in the observations of \cite{Wu2004}, we find the higher metallicity $Z/Z_{\odot}\sim5-9$ is more consistent with the observational results (red/green stars and diamonds in Figure \ref{fig4}).

\subsection{The timescale of line-profile evolution}
To explore the possible variation of the line profile along with the variation of the continuum, we consider the evolution of the dimensionless accretion rate, $\dot{m}$, in our continuum model. Here, we simply assume a power-law evolution of $\dot{m} = \dot{m}_{\rm i} \times (t/3\times10^7 \,{\rm s} + 1)^{\alpha}$, where $\dot{m}_{\rm i}$ is initial dimensionless accretion rate, $\alpha=1$ and -1 describe the rise and decay of the accretion rate respectively. A very large value of $\alpha$ should correspond to a dramatic change in accretion rate or disk spectrum. From the observational point of view, we consider the possible variation of accretion rate on a timescale of several years and test how it will affect the profile of broad lines.

In the left panel of Figure \ref{fig5}, we show the evolution of the line profile for the dimensionless accretion rate increasing from $\dot{m}_{\rm i}=0.02$ to $\dot{m}_{\rm f}=0.2$ (red dashed lines). For comparison, we also present the line profile evolution for a dramatic change from $\dot{m}_{\rm i}=0.02$ to $\dot{m}_{\rm f}=0.2$ directly (blue solid lines), which corresponds to the very fast change of the accretion rate or disk luminosity. It can be found that the two cases are not much different, where the line profile changes from double peak to flat top within several years and then changes to single peak after 30 years. In the right panel of Figure \ref{fig5}, the evolution of line profiles is presented considering the accretion rate change from $\dot{m}_{\rm i}=0.2$ to $\dot{m}_{\rm f}=0.02$. After about 100 years, the single-peaked line profile roughly disappears and new double-peaked line appears in low-accretion-rate state. In this case, the line profile with slow evolution of accretion rate is also roughly consistent with that in dramatic change of accretion rate (from 0.2 to 0.02). The timescale for the variation of the line profile is sensitively related to the initial or final accretion rate. For example, we present the evolution of the line profile for a dramatic change from $\dot{m}_{\rm i}=0.02$ to $\dot{m}_{\rm f}=1$ in Figure \ref{fig6}, which takes only about 1 year for the line profile change from a double peak to a flat top and about 10 years to transition to a single peak. We also simulate a case of a slow change for the very low accretion rate from $\dot{m}_{\rm i}=0.05$ to $\dot{m}_{\rm f}=0.005$ with about 200 days in Figure \ref{fig7}, which may occur in some changing-look AGNs or low-luminosity AGNs. In cases of low Eddington ratios, the line profiles are always double-peaked, with the peak separation becoming wider in a lower accretion state. The timescale for this change is only several years.

\section{Summary and Discussion}\label{discussion}
The wind from the accretion disk is a possible origin for the BLR gas, as observed from the broad lines in AGNs. In this work, we explore the variation of the profile of the broad lines, considering the evolution of the continuum based on the model of dust-driven clouds. We find that the anti-correlation between the double-peak separation and the Eddington ratio is roughly consistent with the observational results for a sample of double-peak emitters in AGNs if the metallicity $Z/Z_{\odot}\gtrsim 5$, which support the moderate or high metallicities in the pc-scale region of AGNs. It takes several years to several decades for the variation of the broad-line profiles when the continuum undergoes strong variation. The timescale strongly depends on the initial or final accretion rate. The evolution of broad emission lines in strong variable AGNs can further test this model and shed light on the BLR physics based on their evolution.

\begin{figure*}[t]
\includegraphics[scale=0.75]{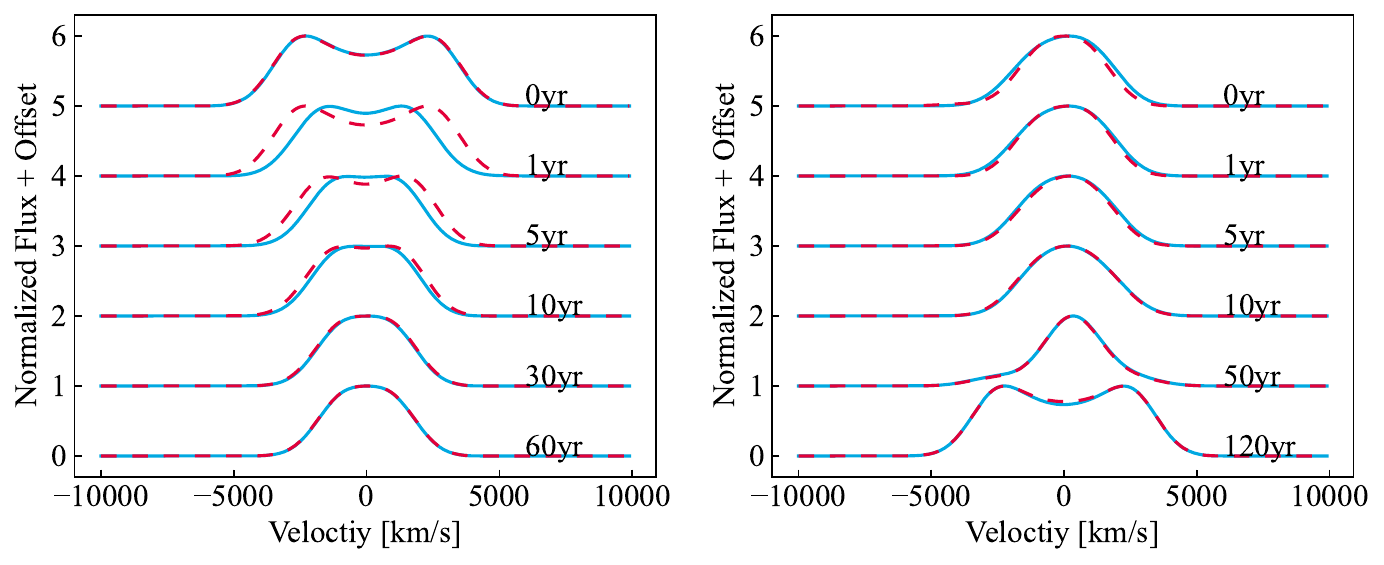}
\caption{The evolution of the broad-line profiles considers the variation of dimensionless accretion rate, where the maximum intensity of each emission line is normalized to one with offsets for different epochs. The left panel shows the results for the dimensionless accretion rate changing from $\dot{m}=0.02$ to $\dot{m}=0.2$, while the right panel presents the evolution with the dimensionless accretion rate changing from $\dot{m}=0.2$ to $\dot{m}=0.02$. In both panels, $M_{\rm BH}=10^8M_{\odot}$, $i=30^{\circ}$, $Z/Z_{\odot}=5$ are adopted. The red dashed lines and blue solid lines represent the slow and dramatic change in the accretion rate, respectively.
\label{fig5}}
\end{figure*}

\begin{figure}[t]
\includegraphics[scale=0.7]{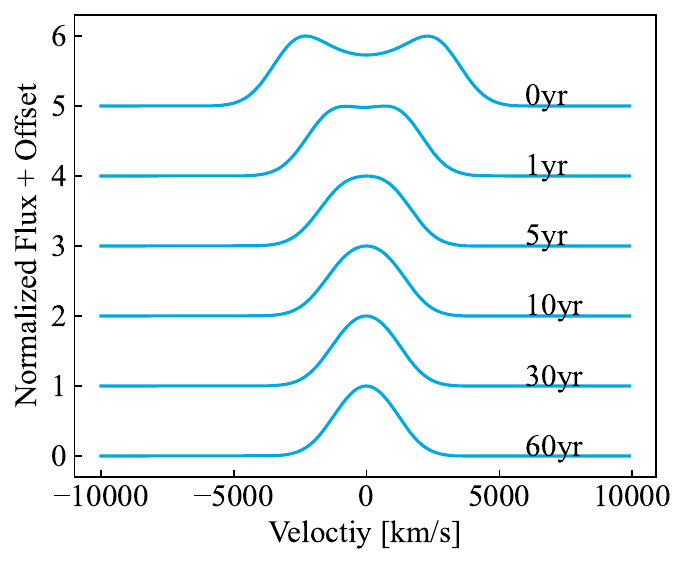}
\caption{Same as Figure 5, but for the variation of the broad line for the accretion rate changing from $\dot{m}=0.02$ to $\dot{m}=1$ with $M_{\rm BH}=10^8M_{\odot}$, $i=30^{\circ}$ and $Z/Z_{\odot}=5$.
\label{fig6}}
\end{figure}
%The structure and kinematics of BLR is 

Based on the assumption of virial motion of gas clouds in the BLR, the masses of SMBHs are widely calculated from the broad-line width and the BLR size (e.g., either the RM method or the empirical correlation of $R_{\rm BLR}-L_{\rm 5100}$ correlation).

Based on the assumption of virial motion of gas clouds in the BLR, the masses of SMBHs are widely calculated from the broad-line width and the BLR size (e.g., either the RM method or the empirical correlation of $R_{\rm BLR}-L_{\rm 5100}$ relation). The unknown BLR structure introduces uncertainties in BH mass measurements derived from BLR gas kinematics. It should be noted that the virial timescale for the clouds in BLR is only several decades or even shorter, where the BLR clouds will collide with the accretion disk. And the BLR will disappear if there is no mechanism for the formation of new clouds. Different BLR formation channels have been proposed in recent decades. \cite{Murray1995} suggested that the line-driven wind produces the broad line feature, which is supported by the high-ionization broad absorption lines as observed in some type I AGNs \citep[e.g.,][]{Weymann1991, Murray1995_2, Gibson2009}. The BLR gas can also be captured from the clumps in the outer torus \citep[][]{Wang2017}. These models can explain some observational features in bright AGNs (e.g., line profile, variability, etc.). However, it cannot explain the double-peak profile, as observed in some nearby low-luminosity AGNs. \cite{Czerny2011} proposed a dust-driven wind model, which can naturally produce single or double-peak broad lines. The origin of the BLR gas is regulated by the accretion process surrounding the central SMBHs, which can also well explain the observed BLR radius-luminosity relation \citep{Czerny2017, Naddaf2021, Naddaf2022}.

At low Eddington ratios, the radiation pressure becomes weaker.  The gas clouds cannot be blown up to a very high position in the dust-driven wind model, leading to a disk-like BLR structure. As a result, it will naturally produce the double-peak broad lines, which is consistent with the observations that the double-peak lines are always found in low-Eddington-ratio sources \citep{Eracleous1994, Storchi2017}. We find the line profile becomes flat-topped or single-peaked when $\dot{m}>0.1$ due to the higher radiation pressure, which is roughly consistent with the observations \citep{Wu2022}. In this work, we explore the relation between the peak separation and other parameters \citep[e.g., inclination, BH mass, accretion rate, and metallicity, see also][]{Naddaf2021}. As the accretion rate decreases, the dust-driven wind can be formed in the inner accretion disk since the dust grains only sublimate at a temperature higher than a critical value. Therefore, the separation of the double peak becomes wider at the lower accretion rates, leading to the anti-correlation between the peak separation and the Eddington ratio. This is consistent with the observations very well, where \cite{Wu2004} found a very tight anti-correlation for a sample of low-luminosity AGNs. The broad-line profile or peak separation is also regulated by the gas metallicity, which regulates the total cross-section of dust and, consequently, the acceleration force. Considering the BH masses in \cite{Wu2004}, we find most of the double-peak AGNs should have moderate or high metallicities ($Z/Z_{\odot}\gtrsim5$). The high metallicity in the pc-scale material of AGNs is consistent with that derived from the broad-line ratios \citep{Hamann1993, Nagao2006, Xu2018, Wang2022, Fan2023}. The high metallicity in the central region of AGNs is widely explored in recent works, where the outer AGN disk is gravitationally unstable and star formation can be formed in the nuclear region of galaxies \citep[e.g.,][]{Wang2010, Huang2023}. This model can roughly reproduce the metallicity enrichment of AGNs \citep[e.g.,][]{Fan2023}.

\begin{figure}[t]
\includegraphics[scale=0.7]{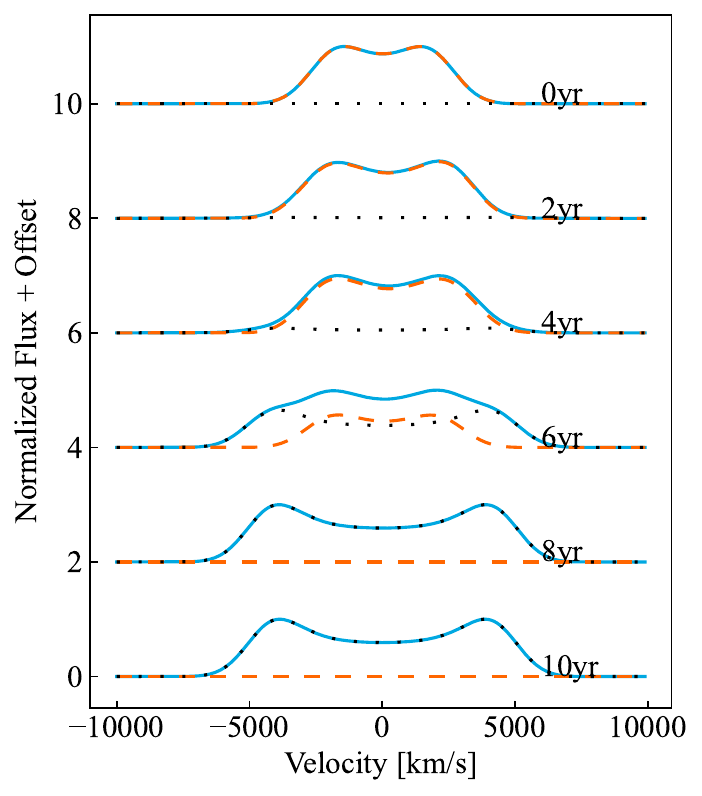}
\caption{Same as Figure 5, but for the line profile variation while the accretion rate changes from 0.05 to 0.005 with $M_{\rm BH}=10^8M_{\odot}$, $i=30^{\circ}$ and $Z/Z_{\odot}=5$. The orange dashed line represents the contribution from the old clouds. The black dotted line represents the contribution from the new clouds. The blue solid line is the total profile.
\label{fig7}}
\end{figure}

Through long-term monitoring in recent years, more and more AGNs with double-peaked emission lines exhibiting strong variability have been reported \citep[e.g.,][]{Lewis2006, Gezari2007, Popovic2023, Lu2024}. This provides a valuable opportunity to explore the BLR physics through the variation of broad-line profiles as the continuum changes. If there is no new BLR clouds formation, the former Kepler-rotation BLR clouds will disappear (i.e., fall back onto the disk) within a timescale of 
\begin{equation}
    t\simeq 15.2 (\frac{\dot{m}}{0.1})^{1/2}(\frac{M_{\rm BH}}{10^8M_{\odot}})^{1/2}(\frac{T_{\rm sub}}{1500{\rm \, K}})^{-2} {\rm \ yrs},
\end{equation}
which is only several years to several decades considering the typical parameter space of AGNs \citep{Czerny2017}. This is consistent with the timescale for the transition from the single peak to double peak or vice visa (see the orange dashed lines in Figure \ref{fig7}). The variation timescale of the broad-line profile is mainly regulated by the disappearance of former equilibrium-state clouds and the launch of new BLR clouds at a later-on accretion state. Many strongly variable AGNs also show strong variation of broad-line profile within several years \citep[e.g., NGC 5548, NGC 4151, Mrk 1018,][]{Peterson2002, Shapovalova2010, Lu2016, Kim2018}. For example, NGC 5548 showed a luminosity decrease from 1998 to 2002 and remained at a low state until 2008, where the peak separation changes from $\sim 2000\,{\rm km/s}$ in 1998 to $\sim 4000\,{\rm km/s}$ in 2004 \citep{Shapovalova2004, Li2016}. This is roughly consistent with our results (e.g., Figure \ref{fig7}) if considering the smaller BH mass $M_{\rm BH} = 5\times10^7M_{\odot}$ \citep{Bentz2015}. The recently reported changing-state quasar of SDSS J125809.31+351943.0 is also believed to have two distinct BLRs by fitting the H$\beta$ line profile with a double-peak profile and a single Gaussian profile \citep{Nagoshi2024}. This is very natural in our model, where different BLR components can co-exist in some periods during the strong variation of the continuum (e.g., Figure \ref{fig5}). More investigations on the broad-line profiles can provide valuable information to constrain the BLR models. For example, the double-peak broad lines also serve as evidence for the existence of the binary black holes \citep[e.g.,][]{Gaskell1983, Tang2009, Kim2018, Terwel2022}. If this is the case, the evolution of the broad-line profile from double-peak to single-peak or vice versa will strongly conflict with the binary SMBHs model.

In calculating the emission line flux from a single cloud, we set a maximum emission-line flux at the critical hydrogen-ionizing photon flux ${\rm log}\Phi_{\rm c}\,({\rm cm^{-2}\, s^{-1}})=18$, which is consistent with the simulation result of \cite{Korista2004} for Balmer lines. It should be noted that different lines may originate from different radii due to the optical depth effect, which is also supported by the RM for different lines \citep{Gaskell2009}. This little difference will not affect the line profiles for low-ionization lines (e.g., H$\alpha$ and H$\beta$). After the dust-driven formation of BLR, we assume the physics of BLR clouds is similar to that in other BLR models when calculating the broad-line intensity. It should be noted that the line intensity itself is not our focus in this work since we mainly consider the relation between the broad-line profile and BLR structure.

\begin{acknowledgments}
The work is supported by the National Natural Science Foundation of China (grants 12233007 and U1931203) and the science research grants from the China Manned Space Project (No. CMS-CSST-2021-A06). We thank M.-H. Naddaf for discussion and data supplement. The authors acknowledge Beijng PARATERA Tech CO., Ltd. for providing HPC resources that have contributed to the results reported within this paper.
\end{acknowledgments}


\vspace{5mm}
% \facilities{}


\bibliography{sample631}{}
\bibliographystyle{aasjournal}

\end{document}